\documentclass[journal]{IEEEtran}

\usepackage{graphics, array, url, hyperref, epsfig, amsmath, amssymb, graphicx}
\usepackage{multirow}
\usepackage{subfigure,color}

\begin{document}

\title{Towards Edge-assisted Internet of Things: From Security and Efficiency Perspectives}

\author{Jianbing Ni, \emph{Member, IEEE}, Xiaodong Lin, \emph{Fellow, IEEE,} Xuemin (Sherman) Shen, \emph{Fellow, IEEE}
\thanks{Jianbing Ni and Xuemin (Sherman) Shen are
 with Department of Electrical and Computer Engineering, University of Waterloo, Waterloo, Ontario, Canada N2L 3G1, email: \{j25ni, sshen\}@uwaterloo.ca.
 }
  \thanks{Xiaodong Lin is with School of Computer Science, University of Guelph, Guelph, Ontario, Canada N1G 2W1, email: xlin08@uoguelph.ca.} \thanks{Corresponding Author: Xiaodong Lin.}
  }


\maketitle

\begin{abstract}
As we are moving towards the Internet of Things (IoT) era, the number of connected physical devices is increasing at a rapid pace. Mobile edge computing is emerging to handle the sheer volume of produced data and reach the latency demand of computation-intensive IoT applications. Although the advance of mobile edge computing on service latency is studied solidly, security and efficiency on data usage in mobile edge computing have not been clearly identified. In this article, we examine the architecture of mobile edge computing and explore the potentials of utilizing mobile edge computing to enhance data analysis for IoT applications, while achieving data security and computational efficiency. Specifically, we first introduce the overall architecture and several promising edge-assisted IoT applications. We then study the security, privacy and efficiency challenges in data processing for mobile edge computing, and discuss the opportunities to enhance data security and improve computational efficiency with the assistance of edge computing, including secure data aggregation, secure data deduplication and secure computational offloading. Finally, several interesting directions on edge-empowered data analysis are presented for future research.


\end{abstract}

\section{Introduction}\label{sec1}
A tremendous number of physical objects embedded with sensors, electronics and actuators are able to exchange information with each others through heterogeneous networks. This phenomenon has culminated in the concept of Internet of Things (IoT) \cite{Gubbia13}. It has provided promising solutions to build resource-hungry and computation-intensive services and applications in various domains, such as smart city, remote e-healthcare, intelligent transportation, industrial automation, autonomous driving and disaster response. This innovation that facilitates interactions among ``things" and human brings new opportunities to transform our society and improve our life.

With the expansion of connected objects, large amounts of data are generated from diverse services and applications, ending up with the overwhelming pressure on data storage, communication and usage. Cisco has predicted that the devices connected to the Internet will generate 507.5 zettabytes per year by 2019. Moving these data from devices to the cloud for storage and analysis would possess a vast number of expensive bandwidth. 45\% of created data would be stored and processed at the edge of networks, or upon close to devices. Moreover, all things generate data constantly, which must be analyzed rapidly to satisfy application requirements. For example, self-driving vehicles are required to process the data collected by on-board sensors, e.g., cameras, radars and lasers, to make real-time decisions. As a result, centralized data storage and analysis cannot afford the demands of data-intensive services or latency-sensitive applications.

Mobile edge computing (MEC) \cite{Sun16} has been envisioned as an enabling technology to support data-driven services and local IoT applications by pushing computing, storage and networking resources from the cloud to the edge of networks. The evolution towards conventional communication networks significantly improves the capability of connections and the utility of network resources between end devices and network ``core", including data centers, cloud and long-term evolution (LTE) core network. With the decentralized resources, fog nodes, e.g., macro/small cell base stations and Wi-Fi hotspots, collaboratively perform a substantial amount of communication, control, storage and management to improve the efficiency of data collection, transmission and analysis, bringing appealing conveniences to IoT applications, including high bandwidth, low latency, location awareness and real-time service access. MEC moves the on-demand latency-sensitive IoT applications proximate to end-users for rapid response and pre-process data produced by devices to reduce the burden on data transmission and analysis for data-intensive services.

Currently, the advance on service delay has been properly investigated in the literature \cite{Sarkar16}, which demonstrates that fog nodes decrease the overall service delay, as the number of IoT applications requesting for real-time service increases. {The approaches of enhancing local data processing using edge computing are also investigated in IoT services. For example, fog computing based content-aware filtering was proposed to enable the fog nodes to perform content filtering for identifying malicious packages in information centric social networks \cite{Dong17}.} As intermediate nodes between core networks and IoT devices, fog nodes are able to pre-process both up-link and down-link data to reduce communication overhead and assist to perform computing tasks offloaded by service providers or IoT devices. In this way, computational and communication efficiency is significantly improved. However, as a powerful intermediate responsible to provide low-latency services and pre-process two-way exchanged data, a fog node obtains all exchanged information unconsciously, resulting in confidentiality corruption for both service providers and devices. While data encryption standards and security protocols, e.g., AES CBC, RFC 5246 and IEEE SA-1735, can achieve end-to-end data confidentiality, they also constrict the pre-processing on the encrypted data for fog nodes and may cause large communication overhead due to the increase of data size.


{In this article, we explore the potentials of utilizing mobile edge computing to enhance data processing in IoT applications, while improving data confidentiality and computational efficiency simultaneously. Specifically, through examining the architecture of MEC and several promising MEC-assisted IoT applications, we demonstrate the challenges on protecting security and privacy, while achieving high efficiency on computation and communication in MEC-assisted IoT applications. Then, the opportunities to enhance the security and efficiency on data processing using edge computing are explored in data-intensive IoT applications, including secure data aggregation, secure data deduplication and secure aided computation, and our insights on achieving secure and efficient data analysis via fog nodes are demonstrated. Therefore, different from \cite{Sun16,Shi16}, we mainly focus on the approaches of exploiting edge computing to benefit data analysis from the perspectives of security and efficiency. Finally, we discuss some interesting and promising research directions to further exploit fog nodes at the network edge to enhance security and efficiency in data-intensive IoT applications.}

\section{Mobile Edge Computing in IoT}
In this section, we introduce the architecture of MEC in IoT and discuss several MEC-assisted IoT applications.
\subsection{Architecture of MEC in IoT}
The systematic and adaptive integration of MEC and traditional access networks expands network connectivity and capability to support nearly omnipotent and ubiquitous network services and IoT applications. By utilizing the network sources at the network edge, fog nodes act as the intermediates to localize IoT services and data storage, and connect the upper cloud layer and the button device layer, as shown in Fig. 1.

\begin{figure}
\centerline{\includegraphics[width=0.5\textwidth]{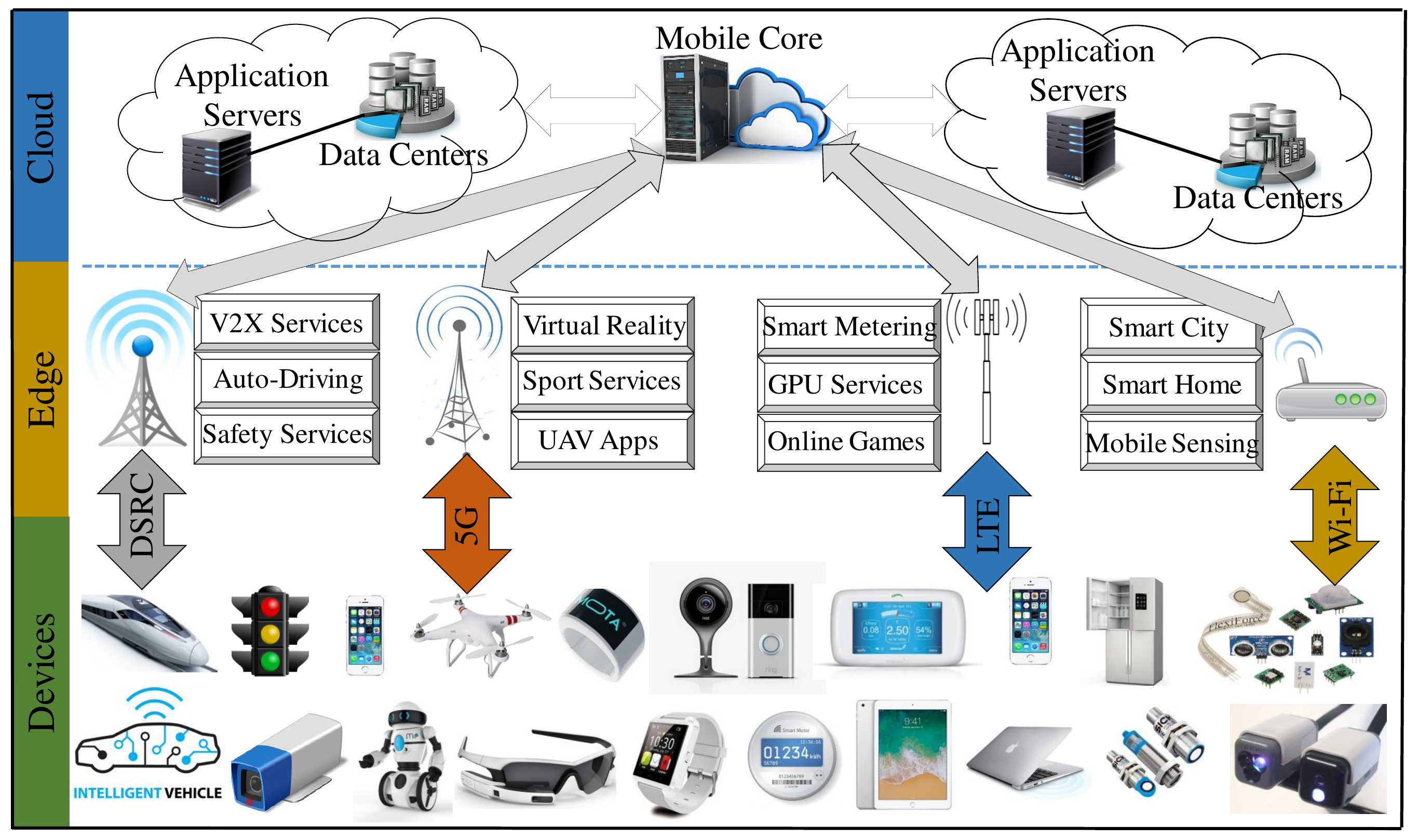}}
\caption{Architecture of MEC for IoT Applications.}
\label{fig:one}
\end{figure}

Cloud--The cloud can be divided into two components. One is mobile core network that provides mobile network connection service to devices, and the other is IP networks, which offer various IoT applications for devices. The mobile core network (e.g., Evolved Packet Core in 4G) is connected with the access networks and IP network, i.e., the Internet. In IP networks, the data centers are responsible to provide outsourced data storage services to devices. The application servers tackle the access requests from devices and utilize rich network, computing and storage resources to offer data-intensive IoT services.

Edge--The edge layer, a network of fog nodes, consists of a large number of macro/small cell base stations and Wi-Fi hotspots. Not only can they provide network connections between devices and core network, but also utilize computing and storage resources to offer various local services and support different IoT applications. Through the recent radio access technologies, e.g., 5G, LTE, mmWave, DSRC, and Wi-Fi, fog nodes realize up-link and down-link data transmissions with pre-precessing to reduce the communication overhead for data-intensive services. Intuitively, fog nodes are extended from traditional base stations and hotspots in access networks with computing and storage capabilities to achieve data pre-processing and caching for IoT applications and enable low-latency services to devices.

Devices--A large variety of devices are connected with mobile ``core" to access various IoT applications. There are two types of IoT devices. The mobile devices are carried by their owners, e.g., fitness trackers, wearable cameras, smart clothes, smart phones, smart watches, smart
glasses and vehicles. The fixed devices include environmental sensors and RFID tags that are pre-deployed at certain areas or on particular products. Embedded with a series of sensors, smart devices are capable of collecting their interested data from environment, delivering produced data to IoT servers through the relay of fog nodes, and accessing latency-sensitive services empowered by fog nodes or remote IoT applications offered by application servers.

Following the architecture shown in Fig. 1, a cloud-fog framework \cite{Yang17} has been proposed to integrate radio access network with MEC by utilizing the software-defined network approach to catalyze the cloud-fog interoperation for reducing service latency and improve data throughput in mobile network. The possibility of empowering integrated MEC with Wifi has been studied \cite{Rimal17} in future 5G networks, aiming at enhancing the quality of services and improving the utility of network resources.

\subsection{Applications of Edge-Assisted IoT}
We briefly introduce several promising MEC-assisted IoT applications, including industrial IoT, autonomous driving and 5G network, which depend on the produced data from devices for service optimization via local data precessing.

\begin{figure*}
\centering
\subfigure[{Industrial IoT}]{
\label{Fig61}
\includegraphics[width=0.325\textwidth]{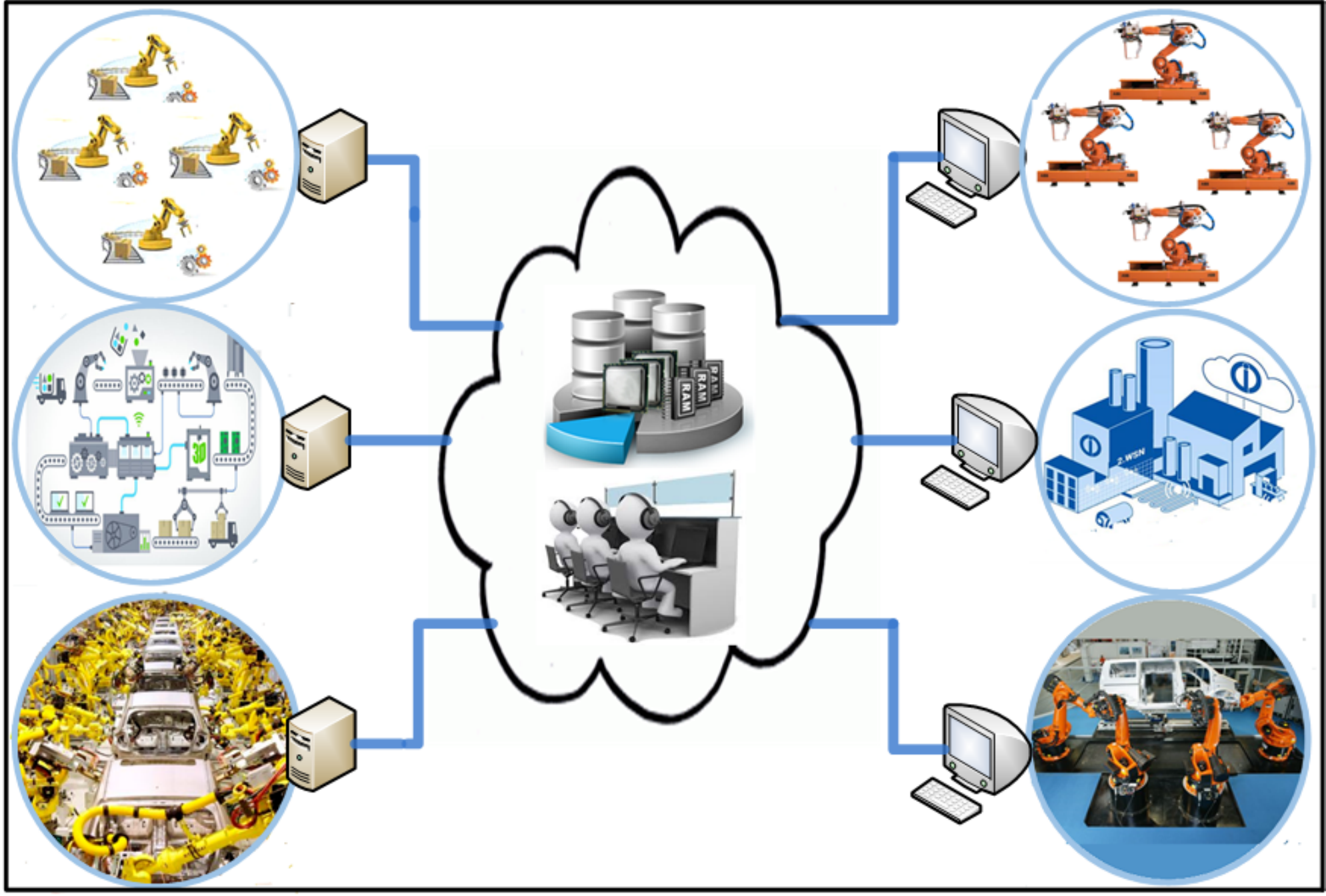}}
\subfigure[Autonomous Driving]{
\label{Fig62}
\includegraphics[width=0.32\textwidth]{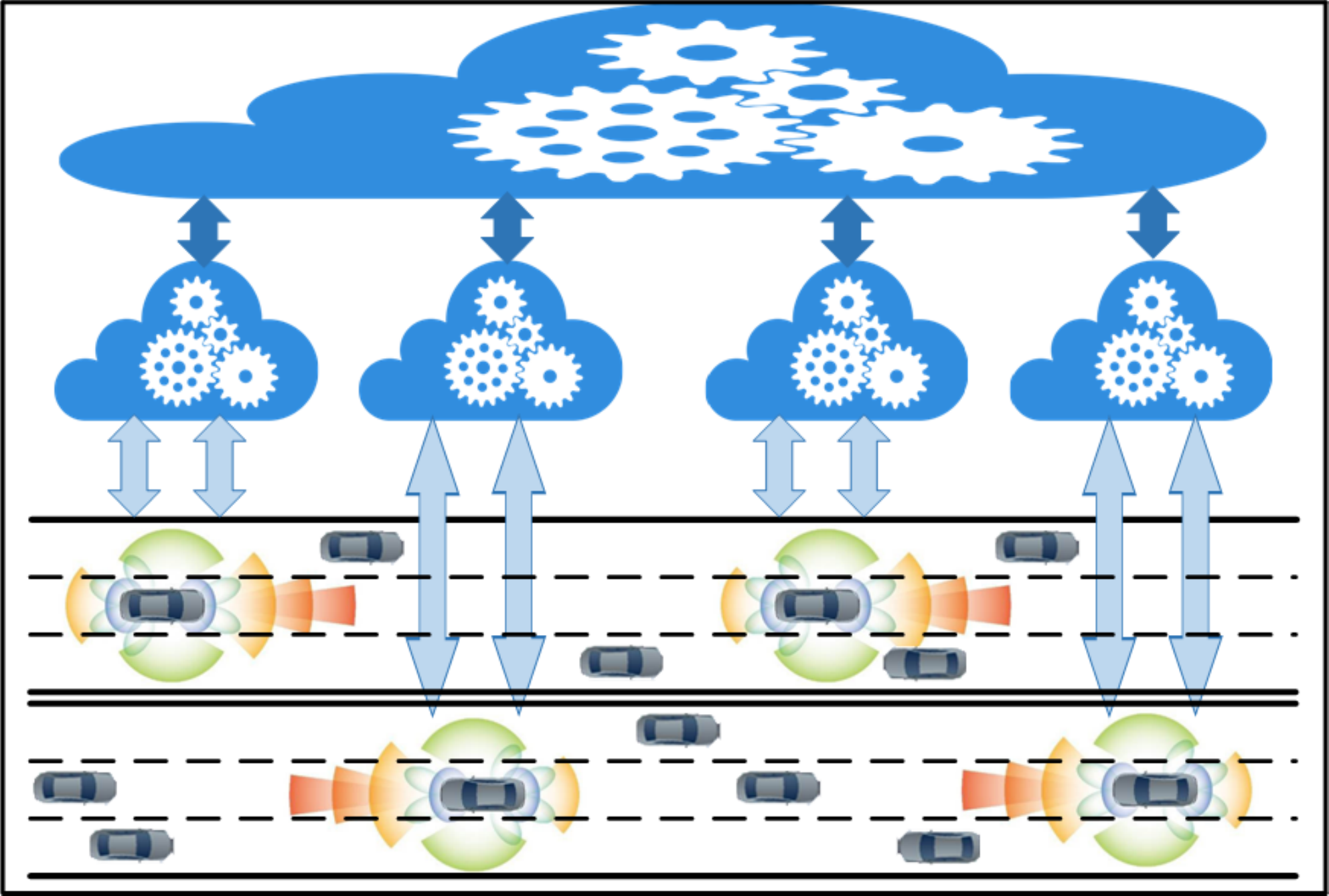}}
\subfigure[5G Networks]{
\label{Fig63}
\includegraphics[width=0.32\textwidth]{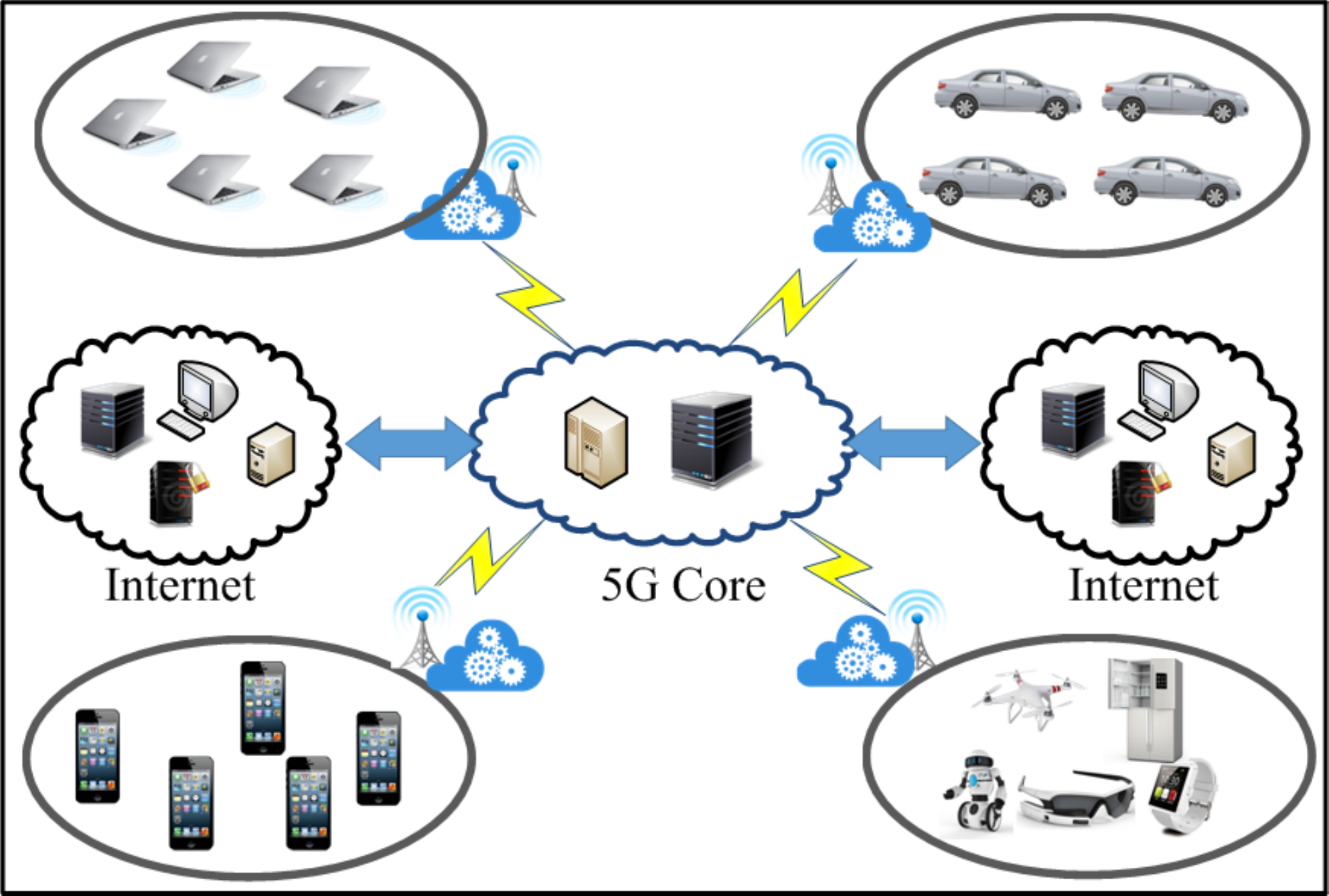}}
\caption{Applications of Edge-assisted IoT}
\label{Fig6}
\end{figure*}

\subsubsection{Industrial IoT}
Industrial IoT offers the capability to monitor a production line in real time using the collected data from a myriad of sensors. Data ingestion is of importance to ensure safe production, improve industrial productivity and reduce production cost. For example, the data coming from a vibration sensor on an excavator's pump might discover mechanical vibrations greater than a safety level. To maintain the produced data, cloud computing has been deployed for data storage requirements. However, cloud-based control introduces an intolerable latency in control response. MEC is gaining popularity that resolves the concerns over cloud storage. By distributing computation, control and storage closer to sensors, edge computing dramatically fasters processing time and decreases network overhead. A local controller (i.e., a fog node) is built on each production line to manage production locally and report a summary to the cloud, rather than all collected data. Therefore, the integration of MEC is significant in industrial IoT due to the appealing potential benefits.

\subsubsection{Autonomous Driving}
Self-driving vehicles--like Baidu's, Uber's and Google's driverless cars--collect a variety of traffic and environmental data from their surrounding areas to constantly aware their locations and make decisions. Each driverless car is outfitted with a myriad of sensors to sense information about current position, pedestrians approaching, vehicle proximity and traffic lights for vehicle monitoring. In addition, autonomous vehicles need to acquire driving routes of other vehicles on roads to analyze driving environments, and thereby optimize their decisions and improve driving safety. In this case, the latency is an essential criteria to achieve autonomous driving, such that MEC is involved to play an essential role for the healthy development of self-driving vehicles. Fog nodes, such as the resource-enriched RSUs or base stations, continuously obtain traffic data from vehicles and tirelessly analyze road conditions in extreme weather, intelligently identify telltale signs and assist vehicles to make correct decisions to respond urgent situations. 

\subsubsection{5G Networks}
5G aims at higher capability than 4G network with the purpose to support a higher density of mobile users and allow massive machine-type communication, enhanced mobile broadband and ultra-reliable low latency communication. 
In doing so, the current radio access networks have been upgraded by many advanced technologies, one of which is MEC. As a key technology enabler in 5G, MEC extends the current 4G networks with new services and business models. By pushing the computing and storage resources from the core network to base stations, MEC-assisted 5G enables various services and applications with the properties of proximity, ultra-low latency, high bandwidth and location awareness. Also, MEC allows the mobile users to real-time access to radio networks and context information. Therefore, MEC advances 4G networks into a programming world to ensure highly efficient network operation, low-latency service delivery, ultimate personal experience and new business opportunities.

\section{Security, Privacy and Efficiency Challenges}
Despite the broad applications of MEC, the security and privacy issues in MEC-assisted application are serious, which are inherently conflict with the computational and communication efficiency. In this session, we point out the security and privacy threats and efficiency challenges to clarify the conflict.

\subsection{Security Threats}
In the data-intensive IoT applications, fog nodes perform the tasks of data pre-precessing to reduce latency and improve network throughput based on their computing, networking and storage resources. It is widely recognized that MEC is a more secure paradigm than the cloud computing due to the property of localization. In general, local data storage decreases the dependency on Internet connections, and local data exchange reduces the possibility of data exposure. Unfortunately, MEC still faces with a variety of cyber attacks because of the limited resources and the drawbacks in secure communication protocols, e.g., WPA2 and TLS1.0. The security threats confronted by MEC are equally serious with cloud-based IoT environments. We discuss three kinds of attacks as examples to demonstrate the security threats in MEC.

{\sf DDoS Attacks}: Due to the limitation on computing, networking and storage resources on a single fog node, it is pretty vulnerable to the distributed denial-of-service (DDoS) attacks, in which an attacker under the control of millions of devices floods the target fog node with superfluous data requests, as a result, this fog node does not have sufficient resources to provide normal services to its intended users. In addition, the lack of security protection of IoT devices intensifies the possibility of DDoS attacks, as it is possible to compromise a large number of IoT devices to build a Botnet for DDoS attacks.

{\sf Corruption Attacks}: {In MEC-assisted IoT applications, fog nodes collect data produced by devices and utilize them to provide local services, such as location-based services, and then transmit the data to the cloud for permanent storage and data analysis. For example, mobile advertising enables brokers to discover the mobile users in a certain area for posting advertisements via the local server, and analyze the conversion and the return on investment on the cloud. However, the rogue or compromised fog nodes may corrupt the produced data or analysis results. }

{\sf Security Threats from Devices}: Nowadays, smart devices are coming under increasing attacks, hard to immune for general users. Hackers are stepping up their cyber attacks on mobile devices with new weapons, including mobile Botnets, ransomware and IoT malware. Over 1.5 million incidents caused by mobile malware were detected by McAfee Labs in the first three months of 2017. These threats not only introduce huge security concerns towards users, but also bring serious security vulnerabilities to the IoT applications on these infected devices, leading to data leakage, data corruption or application dead.

\subsection{Privacy Threats}
In MEC-assisted applications, the user privacy may be leaked out during data processing. The fog nodes extract users' personal information from the data reported by the devices. Here, we discuss two types of privacy leakage: data privacy leakage and location privacy leakage.

{\sf Data Privacy Leakage}: In IoT applications, the produced data encapsulate various aspects of physical environment, some may be considered sensitive, e.g., personal activities, preferences, health status and industrial design drawing, but other may not, e.g., air pollution index, social events and public information. It is widely recognized that the ownership of all produced data should belong to data owners. However, to explore data utility, these data are shared with others without the permission of their owners. As a result, the data, either sensitive or not, would be captured or accessed by the ineligible entities during data transmission and sharing.

{\sf Location Privacy Leakage}: Due to the location awareness of base stations and Wifi hotspots, a device's position can be extracted based on the location of access points. The limited coverage of access points in MEC increases the possibility of location privacy disclosure. Although this is a coarse-grained location exposure, the activity area of a specific device owner is disclosed from the leaked location. Further, if a device builds connections with multiple fog nodes to access their services, the precise location can be obtained using positioning techniques.

\subsection{Efficiency Challenges}
An increasing number of devices in IoT applications produce massive fine-grained or high-precision data, including audio, video and multi-dimensional data. For example, a smart vehicle produces 1 petabytes data everyday and an engine of Boeing 737 generates 333 gigabytes data per minute. It is impossible to forward all these data directly to the cloud for storage and analysis, which possess a large amount of bandwidth in transmission. How to improve the computational and communication efficiency is dramatically essential for data-intensive applications.

In addition, to prevent data leakage, end-to-end encryption is widely used in secret transmission. Unfortunately, it is inevitable to sacrifice computational efficiency, since extra computation is required to encrypt the data before being sent and decrypt to recover the plaintexts. Further, to prevent data corruption during transmission, digital signatures are appended to ensure data integrity and authentication, which trigger extra communication overhead. In summary, the computational and communication efficiency has to be sacrificed for data security.

\section{Opportunity of Edge Computing}
The discussed security, privacy and efficiency issues impede the success of IoT applications. Luckily, MEC brings great opportunities by leveraging its resources at the network edge to resolve challenges. The security and efficiency of data collection and processing can be significantly improved via edge computing in data-driven IoT applications. We discuss several kinds of edge-assisted data processing to light up this promising research topic.

\subsection{Secure Data Aggregation}
Due to large communication overhead and security threats, it is difficult for the cloud to achieve real-time monitoring and demand response based on the produced data of devices in data-intensive IoT applications with the limited resources on mobile devices. Secure data aggregation is an efficient data compression technique with high security protection. To prevent data privacy leakage, each device first encrypts its produced data using a homomorphic encryption scheme, e.g., Paillier, BGN, ElGamal or BGV cryptosystem, before submitting them to the fog nodes. {After receiving the encrypted data from devices, the fog nodes aggregate them, i.e., to compute the summary or multiplication of the individual data, and upload the aggregated results to the cloud. Since the data are encrypted, the fog node cannot learning any knowledge about the individual data reported by mobile users, and the cloud is able to decrypt to acquire the aggregated results, i.e., the summary or multiplication. The communication overhead from the fog nodes to the cloud can be dramatically decreased as only the aggregated results are transmitted to the cloud, the size of which is equal to binary length of an individual data.}  
With the aided computation of fog nodes, secure data aggregation compresses the ciphertexts of delivered data to improve communication efficiency without satisfying the individual data security.

Ni et al. \cite{NITSG17} further enhances data security and reduces communication overhead by improving secure data aggregation in smart grid. In specific, they propose a privacy-preserving smart metering protocol named P$^2$SM to achieve end-to-end security, data aggregation and signature aggregation for preventing the misbehavior of rogue collectors, i.e., fog nodes. The smart meters' readings are encrypted using homomorphic encryption for data confidentiality, and digital signatures are generated by smart meters based on key-homomorphic signatures to prevent data corruption of fog nodes. According to the homomorphism, the collector aggregates both encrypted readings and individual signatures delivered by the smart meters to save communication bandwidth from the collector to the operation center, and reports the aggregated results to the operation center for real-time demand response. Due to the aggregation of individual signatures, it is impossible for a rogue collector to inject false data, so as to prevent the operation center from making false decisions. In addition, verifiable daily billing is achieved to protect the daily bills of customers against rogue collectors and utilities. Therefore, by utilizing the enhanced secure data aggregation, the computational and communication overhead is significantly reduced with end-to-end security protection on the readings of smart meters. Fig 3(a) demonstrates the communication efficiency of P$^2$SM compared with TLS protocol (T-AES), where AES-256 is leveraged to protect readings and the BLS signature is utilized to achieve data integrity. {Different from the solutions mentioned in \cite{Yan12}, which separately solves the privacy, confidentiality, availability and integrity issues in smart grid, P$^2$SM achieves all these properties simultaneously by solving the conflict between secure data aggregation and verifiable daily billing via MEC.
}

\subsection{Secure Data Deduplication}
In data collection, large amounts of data produced by the devices are reductant. These replicate data consume expensive bandwidth for transmission and space for storage. A straightforward approach for bandwidth saving is to discard the replicate copies on intermediates. Nonetheless, this method discloses the produced data to intermediates, which may contain plenty of sensitive information. Data encryption is popular to avoid data leakage, but the detection of replicate copies for intermediates is infeasible, after the data are encrypted, as all the
data are transformed to be random values. To overcome this dilemma, secure data deduplication has been introduced to allow the intermediates to detect the replicate data from different devices without learning any knowledge about the data. This technique has been widely implemented in cloud storage to discover the identical files outsourced by users and delete the same copies to save storage space. In data collection of IoT applications, because of the existence of replicate data, secure data deduplication can be leveraged to improve communication overhead. Although the architecture of data-driven IoT applications is entirely different from cloud storage, the idea of secure data deduplication can be migrated to the MEC-assisted IoT applications.

Ni et al. \cite{NIGC16} introduces secure data deduplication (SDD) to detect and discard the replicate copies of sensing reports submitted by mobile users in mobile crowdsensing, which is a typical IoT application for distributed data collection. The fog nodes collect sensing reports from mobile users and perform data deduplication on behalf of intermediates. A BLS-based oblivious pseudo-random function (BLS-OPRF) is proposed to achieve the identification of replicate sensing reports for fog nodes and prevent brute-force attacks. By leveraging BLS-OPRF, mobile users encrypt the sensing reports, and allow the fog nodes to check whether any two encrypted sensing reports are identical or not. If yes, the fog nodes delete the replicate one. In particular, to record the contribution on task fulfillment of each mobile user, a key-homomorphic signature scheme is leveraged to allow the mobile user to claim her contribution, and the fog nodes are able to aggregate these signatures from multiple mobile users to further save communication overhead. Therefore, the fog nodes can act as intermediates to detect and delete repeated sensing reports for network bandwidth saving without acquiring any knowledge about sensing data. Fig. 3(b) demonstrates that if the number of mobile users and the percentage of duplicates are large, i.e., the points are located in the red area, the communication overhead of SDD is lower than the traditional approach without deduplication. 

\begin{figure*}
\centering
\subfigure[Communication Overhead Comparison]{
\label{Fig62}
\includegraphics[width=0.32\textwidth]{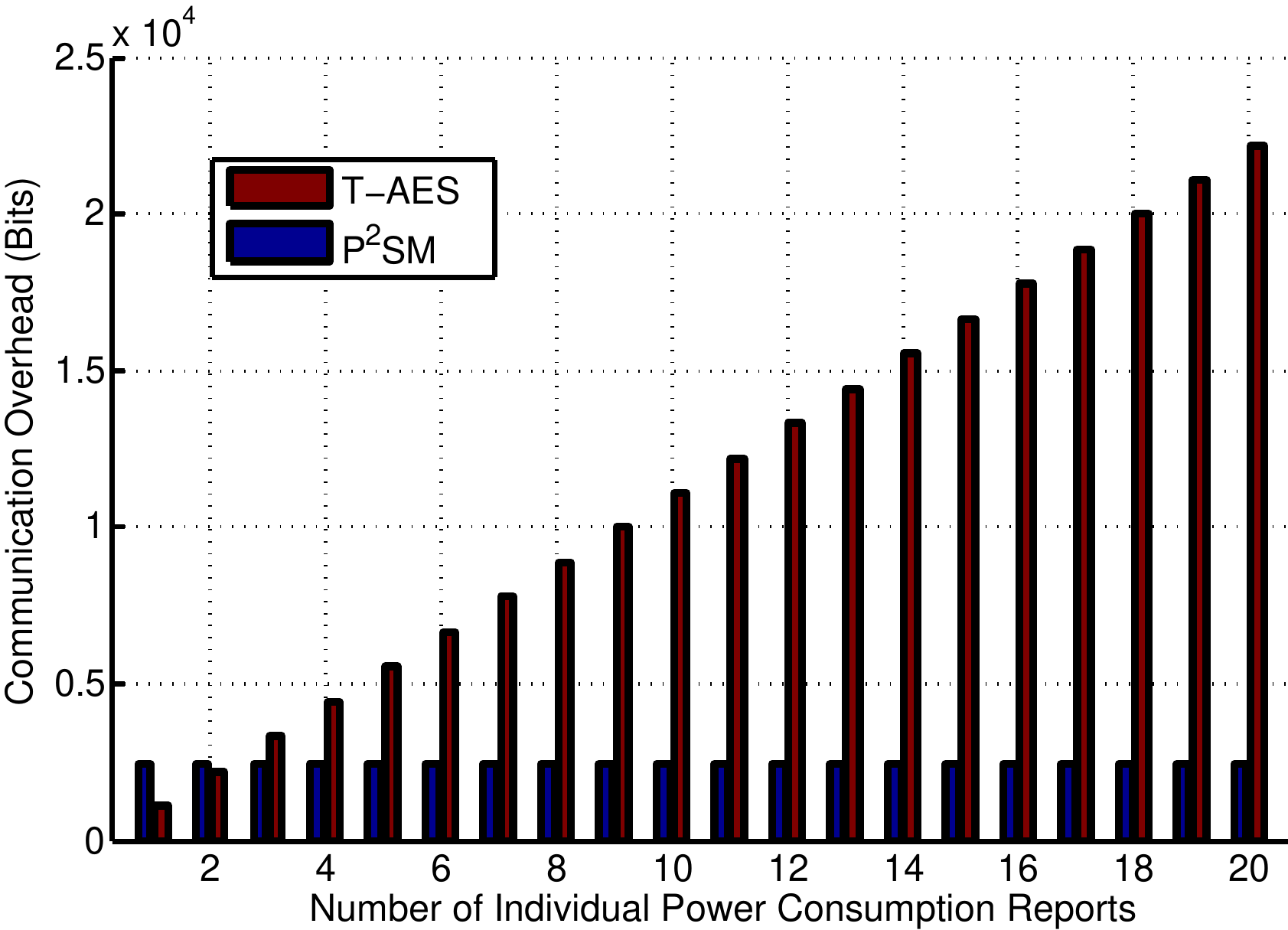}}
\subfigure[Communication Overhead with Deduplication]{
\label{Fig62}
\includegraphics[width=0.32\textwidth]{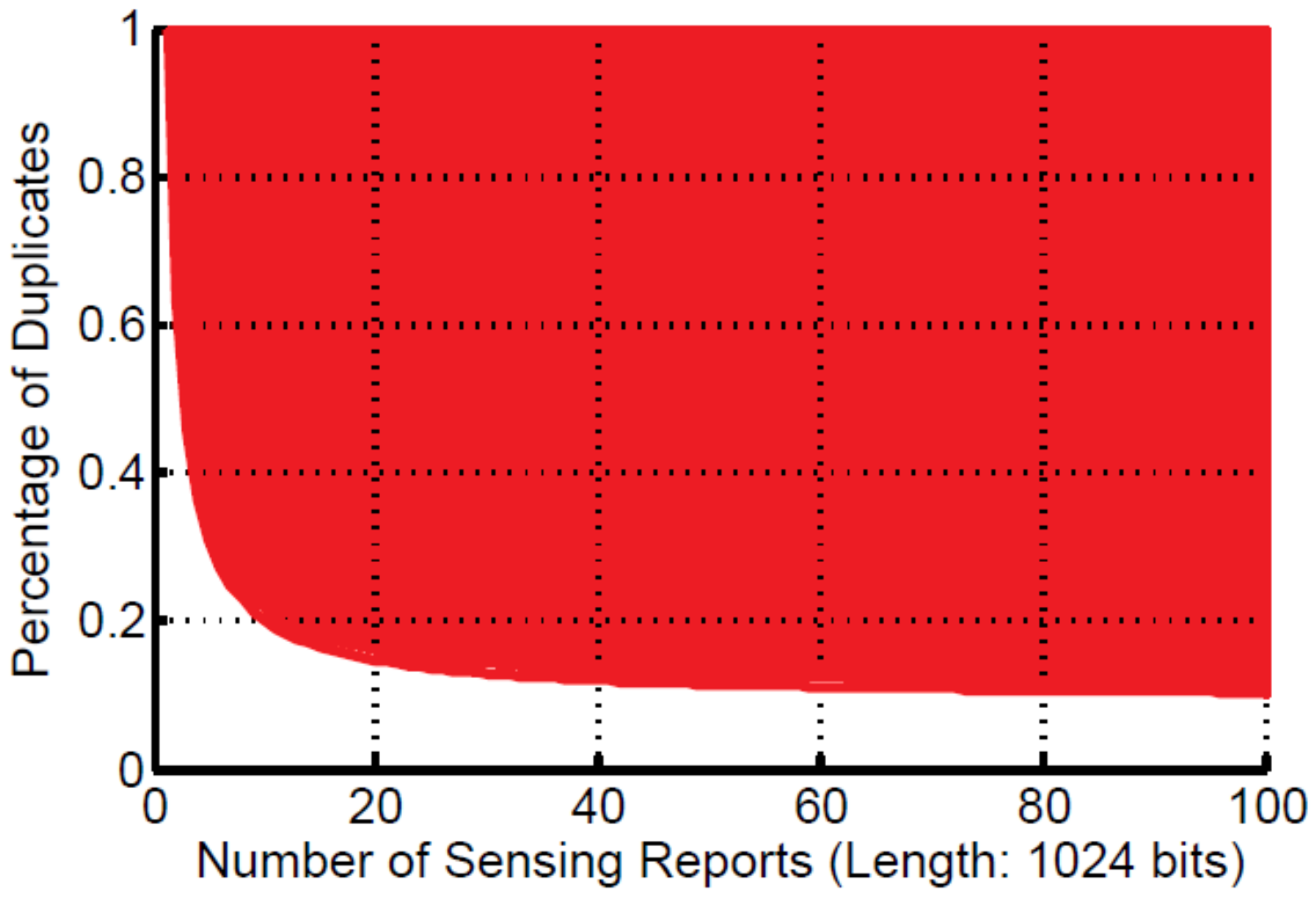}}
\subfigure[Fog-aided Computation in Service Delegation]{
\label{Fig62}
\includegraphics[width=0.32\textwidth]{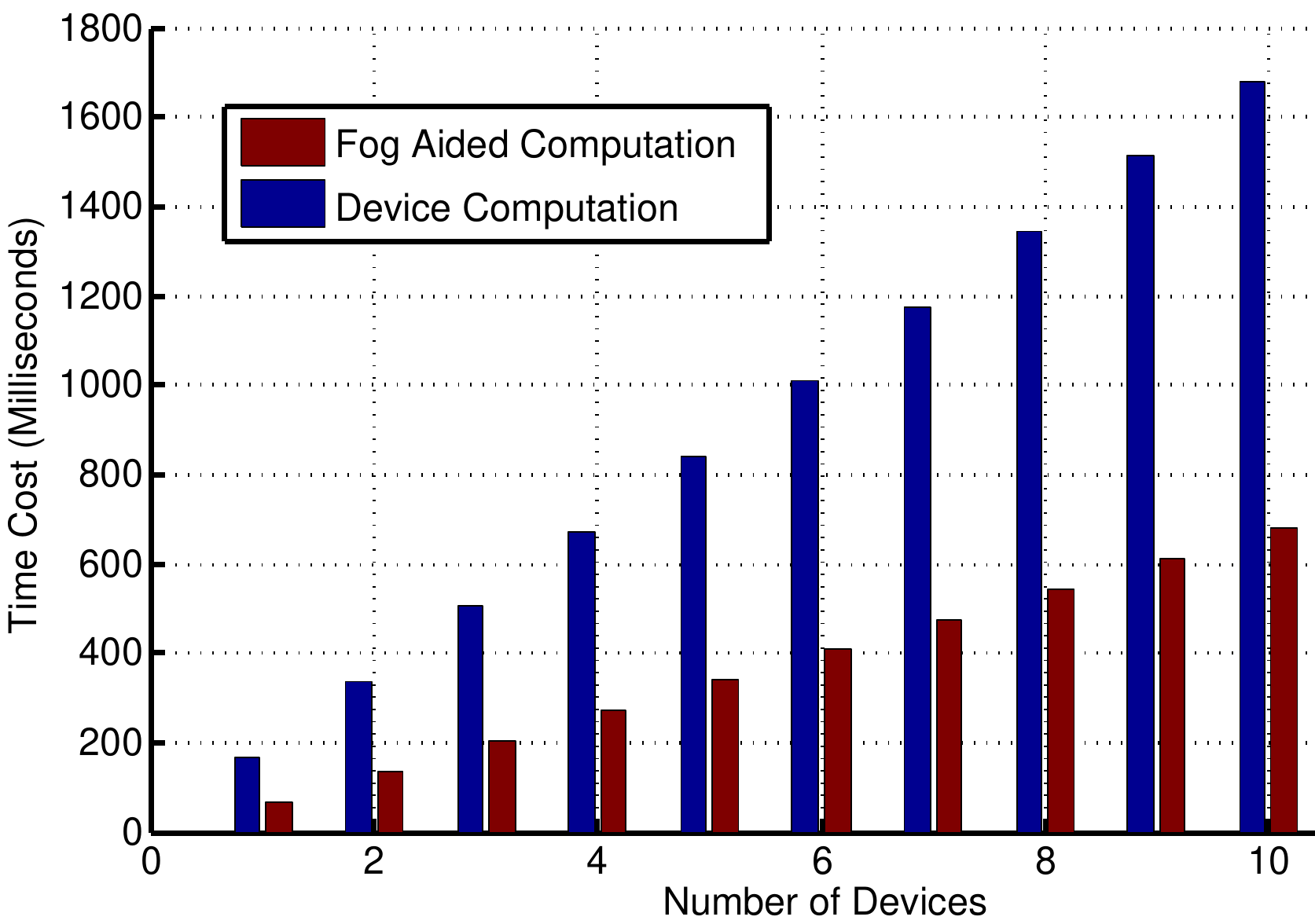}}
\caption{Performance Improvement with MEC Assistance}
\label{Fig6}
\end{figure*}

\subsection{Secure Computational Offloading}
Although the capability of IoT devices becomes more powerful than ever, it is still challenging to support resource-hungry applications due to their constricted resources of energy, storage and computing. Thus, the advanced devices still cannot afford the requirements of data collection, transmission, storage and analysis in large-scale IoT applications. MEC provides an almost perfect solution for resource-limited devices to fulfill the heavy computational tasks \cite{Mach17}. Using the computing and storage resources, some complex operations for devices are offloaded to fog nodes for accomplishment. The migration of computing tasks from IoT devices to fog nodes significantly reduces response delay and computational overhead for devices.
On contrary, fog nodes can also undertake computing tasks offloaded by the cloud.
The fog nodes collaboratively form a femto-cloud to provide computing resources within the access network for satisfying the demands on low latency, frequently access and proximity in IoT services. The two-way computational offloading of MEC is beneficial for both cloud and devices. This computational offloading problem has been formulated to the distributed computing problem based on the real content correlation of data, and the impacts of the computing manner on power consumption has been discussed in \cite{Zhou18}.

Ni et al. \cite{NiJSAC18} proposes an efficient, privacy-preserving service-oriented framework (PPSF) supporting network slicing for 5G-enabled IoT systems with the involvement of fog nodes. On one hand, fog nodes are capable to cache service data frequently visited by users, and select network slices for the coming service packages, so as to reduce service latency and offload computing tasks in slice selection for core networks. A fog node determines the specific network slice for package forwarding based on the match of slice types and service types, while learning nothing about users' interested services. On the other hand, fog nodes perform time-consuming operations, e.g., bilinear pairing, for resource-constrained devices. To migrate computational burden for devices, bilinear pairing operations are executed by fog nodes in signature verification without losing the control on the verification results. Fig. 3(c) shows that the fog-aided computation on bilinear pairing reduces the computational overhead for devices in service delegation of PPSF compared with the method in which the devices preform the operations by themselves. In short, by exploiting the fog nodes at network edges to undertake the offloaded computation and services, the computational overheads of 5G core network and user devices are reduced, and the response latency of IoT services is dramatically decreased for 5G-enabled IoT applications.



\section{Future Research Directions}
MEC offers great opportunities to enhance security and efficiency for data-intensive IoT applications, it also opens several issues for further exploring the advantages of MEC on data processing.

\subsection{Secure Proxy Computation}
IoT devices encrypt and sign the collected data for confidentiality and integrity before uploading them to fog nodes or the cloud. Due to the constrained computing capability and network connectivity, IoT devices may be incapable to carry out the required operations. Proxy computation is an essential candidate to address this issue, in which a proxy is employed to perform the necessary computation under the delegation of devices, such as decryption delegation, signature delegation and re-encryption/re-signature delegation. In MEC-based IoT, a fog node acts as the proxy to support proxy computation. However, the fog node may not be fully honest, indicating that the fog node can abuse the delegation for its own purpose. Therefore, in proxy computation, the main challenge is to design secure proxy computation protocols to prevent the abuse of the delegation.

\subsection{Secure Aided Computation}
Due to the restricted computing capability, it is very challenging for IoT devices to perform complex computation, such as data mining and cryptographic operations, end up with the failure of some vital services. To enable these resources-hungry services, fog nodes play important roles to assist IoT devices for executing complex operations. To achieve aided computation, one straightforward method is to forward all the collected data and the necessity to the fog node. Unfortunately, this approach inevitably discloses all information to the fog node, which may be compromised. Further, in many cases, the secret key is required for performing particular operations, such as signature generation and data decryption. Having the secret key, the fog node can pretend the device to misbehave without being detected. As a result, the security and privacy of the device are entirely lost. Besides, the messages sent or received by the device are under the control of the fog node, and thereby the device may accept a false result. Therefore, it is important to provide sufficient security guarantees in edge-aided computation in IoT applications.

\subsection{Verifiable Computation Offloading}
The fog nodes assist the cloud to perform computing tasks, but whether the results are correct or not is the major concern for the cloud. Verifiable computation enables the cloud to check the correctness of the computing results. As the fog nodes are distributively perform the tasks, the traceability of the misbehaving fog nodes is pretty difficult, in case the returned result is not acceptable. In addition, the device accesses the services offered by local fog nodes with low latency, the correctness of the acquired service results should be verified as well to prevent being cheated. In summary, an efficient verifiable secure computation framework is critical for MEC-assisted computation to guarantee the correctness of computing results.

\subsection{Secure Data Analysis}
Machine learning has recently become extremely popular for its ability to achieve data analytics, which enables to hide insights in a large volume of dataset without being explicitly discovered. Artificial intelligent systems are deployed in a centralized way where a cloud analyzes the training data from individual devices to acquire insights. This model raises serious privacy leakage since the original data may contain sensitive information about individuals. {With the increasing capability of edge devices, such as self-driving cars, drones, and home robots, moving some of functionalities to edge devices is the great trend, which can improve security, privacy, latency and safety. By partitioning execution cross edge devices and the cloud, individuals can locally train their structures and only share a subset of parameters, i.e., the trained model, to keep their respective training set private. However, developing edge-cloud systems is challenging due to the large discrepancy between the capabilities of the cloud and edge devices, and the heterogeneity of edge devices on resources and softwares \cite{Stoica17}. Besides, the collaborative deep learning may be susceptible to disclose the training sets of honest participants \cite{Hitaj17}. Therefore, how to achieve secure data analysis in edge-cloud systems based on distributed and federated deep learning approaches without private training set leakage is still an open problem.
}

\subsection{{Quality of Service (QoS)}}
{The security and efficiency is the inherent trade-off in designing the security protection schemes. The desirable performance metrics include the computational, storage and communication overhead, which roughly evaluate the computing and storage capabilities of devices and end-to-end delay, respectively. However, the key metrics of QoS are rarely mentioned, such as the service availability of edge devices, network throughput, power consumption, and transmission delay. These metrics not only depend on the computing, storage and networking resources of devices, but also vary with the number of connected devices in providing services. The scalability is the fundamental demand to deal with the numerous devices with QoS guarantee in IoT.
}

\section{Conclusions}
In this article, we have investigated the edge-assisted data processing for Internet of Things from security and efficiency perspectives. In specific, we have first presented the overall architecture of MEC-assisted IoT and introduced several promising applications. Then, we have offered a comprehensive overview on the security, privacy and efficiency issues in MEC and discussed the dilemma between security and efficiency in data usage. Further, we have provided the research opportunities to solve the dilemma in MEC-assisted IoT, including secure data aggregation, secure data deduplication and secure computation offloading. Several interesting issues have been identified to shed light on the further research on secure and efficient data analysis in MEC-assisted IoT.

\vfill\eject

\end{document}